\documentclass[10pt]{iopart}
\usepackage{pdf14}
\usepackage{amssymb}
\usepackage[normalem]{ulem}
\usepackage{xcolor}
\usepackage{cancel}
\usepackage{currfile}
\usepackage[colorlinks=true,urlcolor=blue]{hyperref}
\usepackage{graphicx}
\usepackage{textcomp}
\usepackage{gensymb}
\usepackage{siunitx}
\usepackage{lipsum} 
\usepackage{tikz}
\usepackage{footmisc}
\usepackage{slashed}
\usepackage{cite}
\definecolor{TransferRed}{rgb}{0.923,0.11,0.14}
 
\begin{document}
\title{Quantum probe of an on-chip broadband interferometer for quantum microwave photonics}
\paper[Quantum probe of an on-chip broadband interferometer for quantum microwave photonics]{}
\author{$\textrm{P. Eder}^{1,2,3}\footnote{mailto:peter.eder@wmi.badw.de}\textrm{, T. Ramos}^{4,7}\footnote{mailto:t.ramos.delrio@gmail.com}\textrm{, } \textrm{J. Goetz}^{1,2} \textrm{, M. Fischer}^{1,2,3}\textrm{, S. Pogorzalek}^{1,2}$
$\textrm{J. Puertas Martínez}^{5,6}\textrm{, E.P. Menzel}^{1,2}\textrm{, F. Loacker}^{1,2} \textrm{, E. Xie}^{1,2,3}\textrm{, }$
$\textrm{J.J. Garcia-Ripoll}^{4}\textrm{, }\textrm{K.G. Fedorov}^{1,2}\textrm{, } \textrm{A. Marx}^{1}
\textrm{, F. Deppe}^{1,2,3}\textrm{,}$
$\textrm{and R. Gross}^{1, 2, 3}\footnote{mailto:rudolf.gross@wmi.badw.de}$}

\address{$ˆ1$ Walther-Meißner-Institut, Bayerische Akademie der Wissenschaften, Walther-Meißner-Str. 8, 85748 Garching, Germany, }
\address{$ˆ2$ Physik-Departement,  Technische  Universität  München,  James-Franck-Str. 1,  85748  Garching, Germany,}
\address{$ˆ3$ Nanosystems Initiative Munich, Schellingstraße 4, 80799 München, Germany, }
\address{$ˆ4$ Instituto de F\'{i}sica Fundamental, IFF-CSIC, Madrid E-28006, Spain, }
\address{$ˆ5$ Universite Grenoble Alpes, Institut NEEL, F-38000 Grenoble, France, }
\address{$ˆ6$ CNRS, Institut NEEL, F-38000 Grenoble, France}
\address{$ˆ7$ Centro de \'Optica e Informaci\'on Cu\'antica, Facultad de Ciencias, Universidad Mayor, Chile}


\begin{abstract}

Quantum microwave photonics aims at generating, routing, and manipulating propagating quantum microwave fields in the spirit of optical photonics. To this end, the strong nonlinearities of superconducting quantum circuits can be used to either improve or move beyond the implementation of concepts from the optical domain. In this context, the design of a well-controlled broadband environment for the superconducting quantum circuits is a central task. In this work, we place a superconducting transmon qubit in one arm of an on-chip Mach-Zehnder interferometer composed of two superconducting microwave beam splitters. By measuring its relaxation and dephasing rates we use the qubit as a sensitive spectrometer at the quantum level to probe the broadband electromagnetic environment. 
For frequencies near the qubit transition frequency, this environment can be well described by an ensemble of harmonic oscillators coupled to the transmon qubit. 
At low frequencies $\omega\,{\rightarrow}\,0$, we find experimental evidence for colored quasi-static Gaussian noise with a high spectral weight, as it is typical for ensembles of two-level fluctuators. Our work paves the way towards possible applications of propagating microwave photons, such as emulating quantum impurity models or a novel architecture for quantum information processing.

\end{abstract}
\pacs{74.50.+r, 03.67.-a, 85.25.Cp}
\submitto{\SUST}
\maketitle
\ioptwocol
\section{Introduction}
With the advent of superconducting circuits as a key player in the field of quantum science and technology, the quantum properties of the microwave signals emitted from these circuits have become a popular object of study. The related experiments can be divided into two major groups. In one of them, the focus is put on the implementation of continuous-variable quantum protocols\,\cite{Menzel2012, Fedorov2016, Eichler2011, Flurin2012, Wilson2011,GoetzPRL2017}. The other group aims at realizing scattering experiments of microwave photons of a quantum system using either microwave photons in a discrete-variable description or quasi-classical coherent states as probe signals\,\cite{Bozyigit2010, Astafiev2010, Lang2013, HoiPhotonRouter2011, HoiCrossKerr2013,Zhou2008,Lu2014}. This latter set of experiments, to which also this work contributes, is closely related to the concepts of generation, routing, manipulation, and detection. These areas have been actively explored in photonics at optical frequencies and are therefore often referred to as microwave quantum photonics \cite{Fan2016, GU20171}. 

In this work, we are interested in the scattering of an incident microwave field off an artificial atom placed in a broadband, but nevertheless carefully engineered, open quantum circuit. In theory, such a system can be described in the framework of the spin-boson model (SBM)\,\cite{Leggett1987, Haeberlein2015}. In this model description the relaxation and dephasing rates of the qubit are determined by the spectral function of the electromagnetic environment. Therefore, measuring the qubit relaxation and dephasing rates over a wide range of the qubit transition frequency allows one to obtain valuable information on the environment. 
\begin{figure}[!b]
  \includegraphics[scale=1]{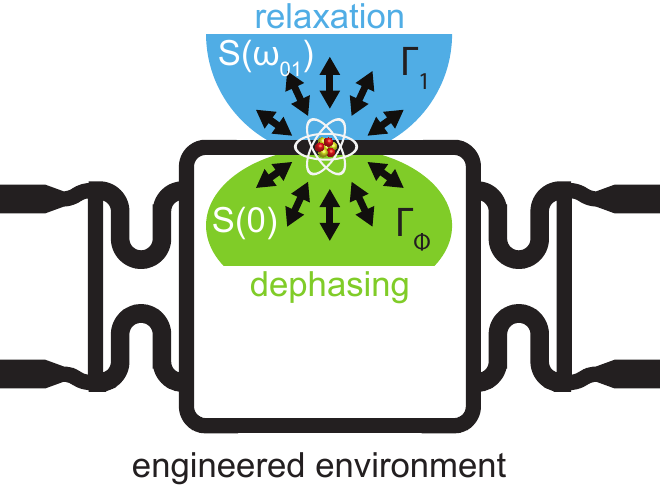}
  \centering
  \caption{An artificial atom realized as a transmon qubit is placed in one arm of a Mach-Zehnder type interferometer (black structure). 
	The decoherence properties of the qubit are determined by its coupling to the environment. We extract the relaxation rate $\Gamma_1$ and the pure dephasing rate $\Gamma_\varphi$ at different qubit transition frequencies $\omega_\textrm{01}$ from microwave transmission experiments. From the derived $\Gamma_1$ values we extract the influence of high-frequency noise (blue) near $\omega_\textrm{01}$ and conclude that the qubit predominantly interacts with the Ohmic bath provided by the transmission line.
At low frequencies ($\omega\,{\rightarrow}\,0$), 1/$f$-noise mainly contributes to the dephasing of the qubit (green). 
}
  \label{fig:story}	
\end{figure}
Getting such information on the environmental spectral function is crucial both for designing quantum technology and studying fundamental properties of quantum coherence.
In particular, one can experimentally determine the microwave environment provided by the implemented circuit and verify, whether or not the actually measured environment coincides with the designed one. Most importantly, identifying the origin of differences between the realized and designed environment provides a tool for deliberate environment engineering.  
This is highly relevant when it comes to the implementation of complex quantum circuits, as required for quantum information processing (QIP) with propagating photons in the spirit of all-optical QIP \cite{KLM2001,Obrien2009,Kok2007,Broome2012}. In addition, one may get valuable information on external sources of decoherence and how to remove them. In this way, quantum coherence, one of the key figures of merit in quantum technology, can potentially be improved.

In our specific experiment, we implement a Mach-Zehnder interferometer based on two on-chip microwave beam splitters and place a transmon qubit in one interferometer arm. The qubit acts as a highly sensitive spectrometer, whose relaxation rates are expected to be dominated by the coupling to the broadband electromagnetic environment provided by the interferometer\,(cf.\,figure\,\ref{fig:story}). The analysis of the experimental results confirms this expectation. Our work makes a major step forward towards the controlled implementation and characterization of complex broadband on-chip circuits required for advanced microwave quantum photonics. 
\begin{figure}[!bh]
  \centering
	\includegraphics[scale=1]{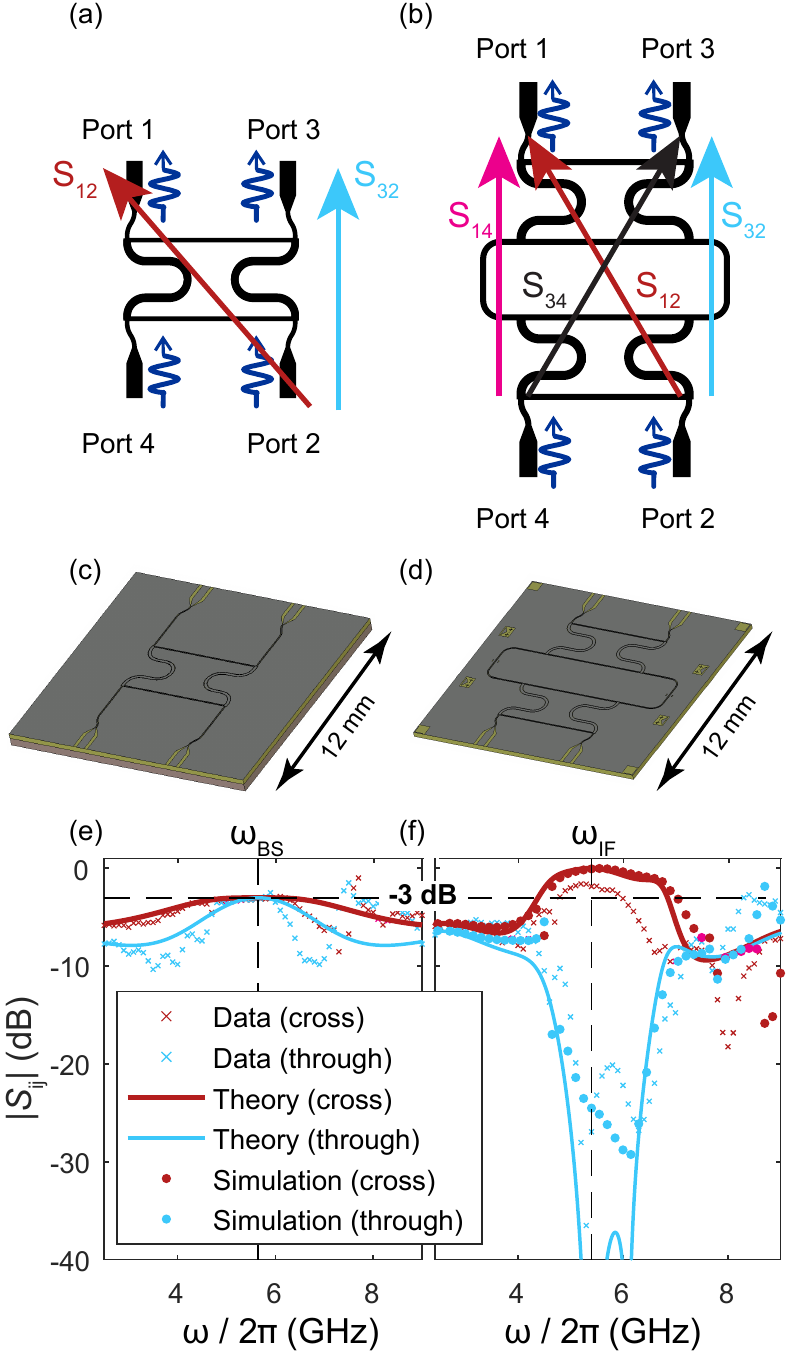}
	\caption{Schematic circuit layout: (a) Beam splitter, (b) Interferometer. 
	Colored arrows mark the transmission $S_{ij}$ for different input and output ports. Cross: $S_{12}$(red) / $S_{34}$(black), through: $S_{32}$(cyan) / $S_{14}$(magenta). The indices $i$ and $j$ identify the output and the input port, respectively.
 (c), (d) Sample layouts of (c) a quadrature hybrid beam splitter and (d) an interferometer.
(e), (f) Measured transmission magnitude (crosses) plotted versus frequency together with the result of the analytical theory (solid lines) and the numerical simulation (full circles) for the beam splitter (left) and the interferometer (right). Cross: $S_{12}$, through: $S_{32}$.
}
		\label{fig:IFBSSchematicWMI}
\end{figure}

This article is structured as follows: First, in section\,\ref{sec:ExpDetails}, we give an overview of the experimental setup and discuss preliminary measurements used to characterize both the interferometer and the transmon qubit via transmission microwave measurements with the help of a vector network analyzer. In section\,\ref{sec:Results}, we present the spectral behavior of the composite qubit-interferometer system and provide a detailed data analysis. In particular, we describe three fundamental regimes of operation and the mathematical model used for analyzing our data.
Fitting the experimental data provides two key quantities characterizing the qubit environment. 
First, the linear increase of the qubit relaxation rate $\Gamma_1$ with increasing qubit transition frequency in the range between $\SI{4}{\giga\hertz}$  and $\SI{7.5}{\giga\hertz}$ clearly shows that the environment provided by the interferometer arm can be well described by an Ohmic spectral density $J(\omega) \propto \omega$ associated with memory-less damping Markovian dynamics. 
Above $7.5\,\si{\giga\hertz}$, we find deviations from this simple behavior, most likely due to weak parasitic resonant modes on the chip. 
Second, we analyze the pure dephasing rate $\Gamma_\varphi$, which indicates the presence of 1/$f$-noise.
Finally, in section\,\ref{sec:summary}, we conclude and give an outlook on future experiments and promising applications based on our platform.

\section{Experimental details}\label{sec:ExpDetails}
In this section, we present details on circuit fabrication, measurement setup, and precharacterization.\\
\noindent\textbf{Fabrication} \textemdash\,Beam splitters and interferometer are fabricated from a $\SI{100}{\nano m}$ thick superconducting Nb-film  sputter-deposited on top of a $\SI{525}{\micro\meter}$ thick Si substrate covered with $\SI{50}{\nano\meter}$ of thermal oxide. 
Both the microwave resonator used for precharacterization and all coplanar waveguide (CPW) structures on the interferometer chip are patterned by optical lithography and reactive ion etching. 
The transmon qubits are fabricated with aluminum technology and shadow evaporation\,\cite{Dolan1977}.

\label{sec:Cryogenics}
\noindent\textbf{Millikelvin setup} \textemdash\,For all measurements involving a qubit, the sample chip is placed inside a copper box and mounted at the base temperature stage ($30\,\si{\milli\kelvin}$) of a dilution refrigerator. 
To minimize the influence of noise leaking to the sample via cabling, 
attenuators thermally anchored at every temperature stage of the refrigerator are used in the input lines.
The output signals are amplified with a chain of cryogenic HEMT and room temperature RF amplifiers.
Two circulators, one in each output line, protect the sample from the amplifier noise. More details on the measurement setup can be found in Appendix A.

\label{sec:PreCharacterization}
\noindent\textbf{Precharacterization}\textemdash\,Before we investigate the coupled qubit-interferometer system, we characterize the two system components independently. 
We start with the characterization of the Mach-Zehnder type microwave interferometer \,\cite{Mach1892} composed of two equivalent quadrature hybrid beam splitters as shown in figure\,\ref{fig:IFBSSchematicWMI}(a) and (c)\,\cite{pozar, Fischer2014,Schneider2014}.

\begin{figure}[!t]
  \centering
	\includegraphics[scale=1]{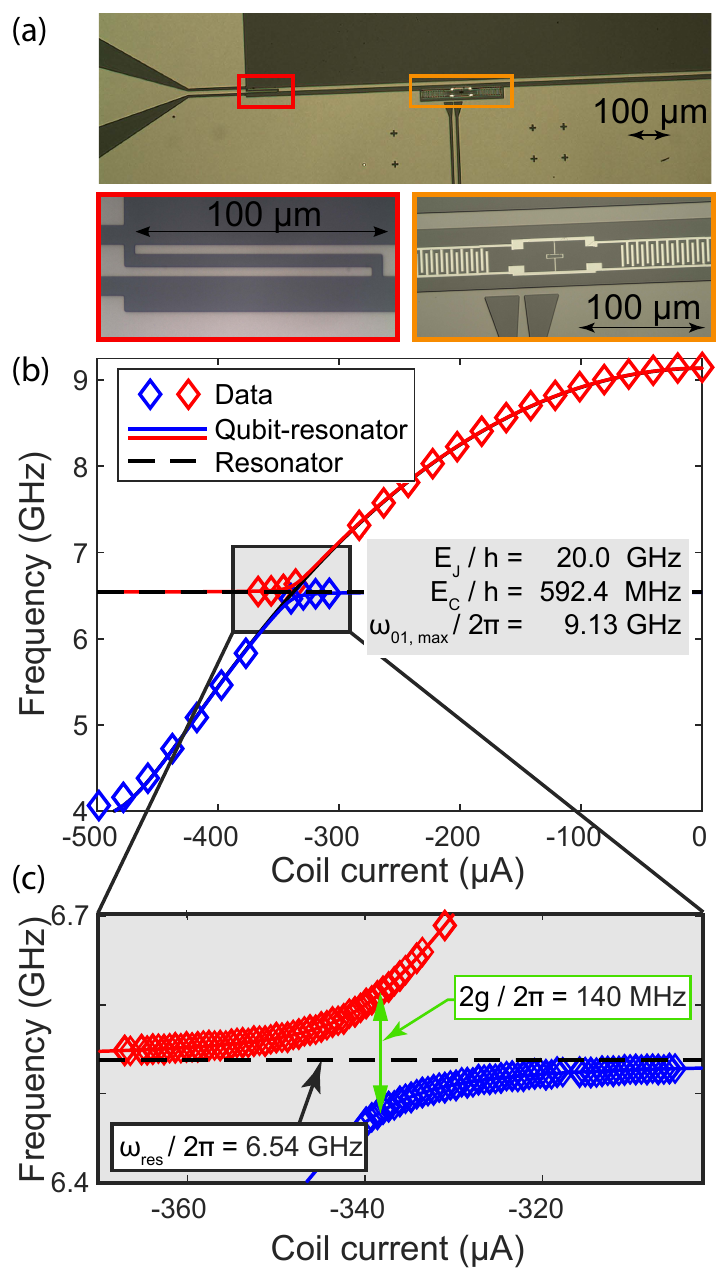}
	\caption{(a) {Optical micrograph of the transmon qubit placed in a $\lambda/4$ coplanar waveguide resonator for precharacterization. The regions marked by the red and orange rectangles contain the coupling capacitors and the transmon qubit, respectively, and are shown on an enlarged scale. }
	(b) Flux dependence of the transmon qubit transition frequency measured by a resonator using two-tone spectroscopy\,\cite{Javier2015}. 
	(c) Zoom-in of region around the avoided crossing. Fitting the data yields the resonance frequency of the $\lambda/4$ resonator $\omega_{\textrm{res}}=\SI{6.54}{\giga\hertz}$, the qubit-resonator
coupling $g/(2\pi)=\SI{70}{\mega\hertz}$ as well as $E_J/h=\SI{20.0}{\giga\hertz}$, $E_C/h=\SI{592.4}{\mega\hertz}$ and $\omega_{\textrm{01,max}}=\SI{9.13}{\giga\hertz}$.}
	\label{fig:Transmon}
\end{figure}
	
We cool down both the beam splitter and the interferometer to $\SI{4.2}{\kelvin}$ in a liquid helium bath cryostat\,\cite{Fischer2014,Schneider2014} and use a vector network analyzer to measure the transmission magnitude $|S_{ij}|$. The indices $i$ and $j$ identify the output and the input port, respectively.
%
%
%
%
%
%

\noindent The measured transmission spectra shown in figure\,\ref{fig:IFBSSchematicWMI}(e) (f) agree well with those obtained from analytical theory\,\cite{pozar} and from finite element simulations\textsuperscript{\footnotemark{}}\footnotetext{\,CST MICROWAVE STUDIO \textsuperscript{\textregistered}, \href{www.cst.com}{www.cst.com}}. 
The interferometer shows a typical isolation of more than $\SI{20}{dB}$ near its designed working 
frequency $\omega_\textrm{IF}/2\pi=5.746\ \si{GHz}$. At $\omega/2\pi=5.9\ \si{GHz}$ we find the maximum isolation of approximately $\SI{30}{dB}$.\\
In order to obtain independent reference data for the qubit parameters and the coupling strength between the qubit and the interferometer, we fabricate a transmon qubit\,\cite{Koch2007} with the same design parameters and place it at the voltage-antinode of a $\lambda/4$ resonator with a resonance frequency $\omega_\textrm{res}/ 2 \pi = \SI{6.54}{\giga\hertz}$. This resonator has a coplanar waveguide design with the same geometry as the transmission line arms connecting the two beam splitters forming the interferometer in the coupled qubit-interferometer system. 
To access the qubit transition frequency far away from the resonator frequency we perform two-tone measurements\,\cite{Schuster2005} using a vector network analyzer and an additional microwave source. The results are shown in figure\,\ref{fig:Transmon}(b) and (c). As expected from the chosen qubit design, the transition frequency of the transmon qubit is tuneable by an applied magnetic flux over the wide frequency range of $\omega_{\textrm{01}}=\SIrange{4}{9}{\giga\hertz}$.  Fitting the data, we derive charging energy $E_C/h=\SI{592.4}{\mega\hertz}$, a Josephson coupling energy $E_J/h=\SI{20}{\giga\hertz}$, and a qubit-resonator coupling $g/(2\pi)=\SI{70}{\mega\hertz}$.

\begin{figure}[!b]
  \centering
	\includegraphics[scale=1]{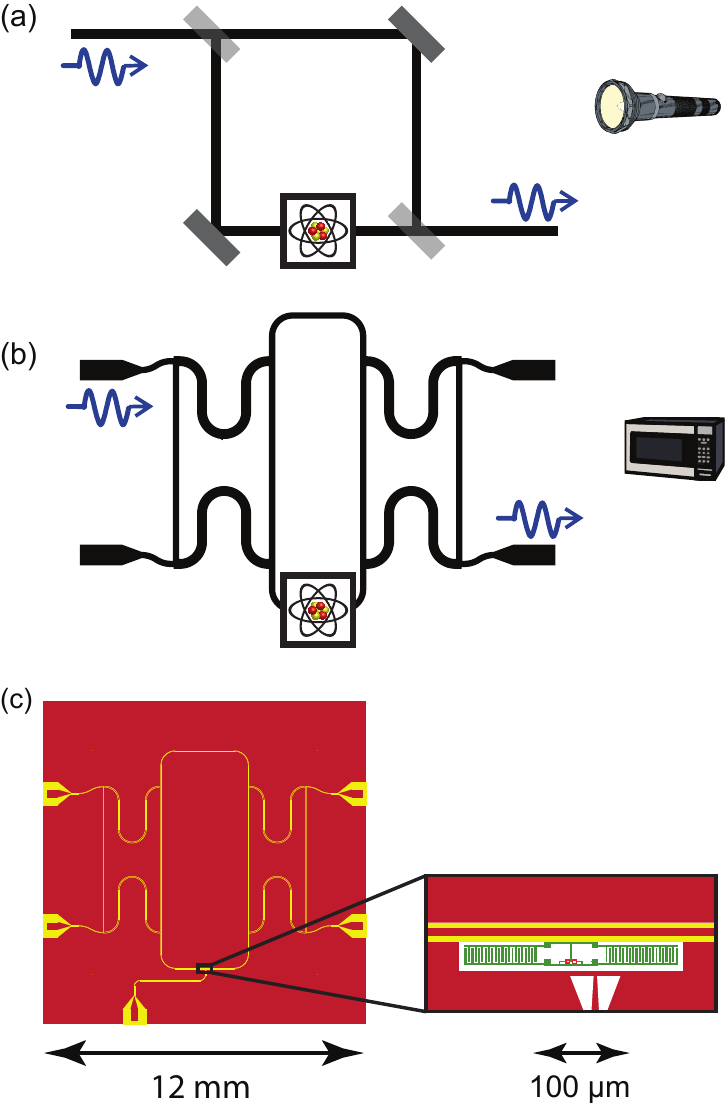}
	\caption{Schematics of (a) an optical and (b) a microwave Mach-Zehnder interferometer with an artificial atom placed in one interferometer arm. 
	(c) Layout of the composite chip, which is discussed in detail in section\,\ref{sec:Results}.
	    The close-up on the right shows the transmon qubit with a tuneable Josephson junction
formed by a dc-SQUID. The positions of the two SQUID junctions are marked with red rectangles.}
	\label{fig:circuitOptical}
\end{figure}



%
%
%
%
%
%
%
%
%


\noindent\textbf{Transmon in interferometer} \textemdash\, The sample under study in section\,\ref{sec:Results} of this work consists of a transmon qubit placed in one arm of a Mach-Zehnder type interferometer. Both the geometry of the 
transmission lines forming the interferometer arms as well as the design and fabrication parameters of the transmon qubit are similar to those of the resonator and transmon qubit used for precharacterization. Details on the sample layout can be found in figure\,\ref{fig:circuitOptical}.\\

\section{Results and Discussion}\label{sec:Results}
In this section, we discuss the properties of the qubit-in-interferometer system. First, we qualitatively describe the transmission spectrum (section\,\ref{sec:regimes}) and then develop a quantitative model based on transfer-matrices (section \ref{sec:model}). We then fit the measurement data to this model in order to extract the qubit transition frequency $\omega_\textrm{01}$, the relaxation rate $\Gamma_1$, and the dephasing rate $\Gamma_\varphi$. We use these results to confirm the validity of our model. 
In section\,\ref{sec:ohmic}, we use the qubit as a sensitive spectrometer of its environment and take a closer look at its decoherence properties.
By analyzing the derived $\Gamma_1$ values for a wide range of qubit transition frequencies we can conclude that the qubit predominantly interacts with an Ohmic bath provided by the broadband transmission line. Additionally, we find evidence for a broad parasitic mode at a frequency above $8\,\si{\giga\hertz}$.
Finally, exploiting $\Gamma_\varphi$ measurements, we discuss dephasing noise at frequencies much smaller than $\omega_\textrm{01}$ in the framework of the Ornstein-Uhlenbeck model\,\cite{vanKampen2003,Ramos2018}.

\subsection{Qualitative analysis}\label{sec:regimes}
\begin{figure}[!t]
 \centering
        \includegraphics[scale=1]{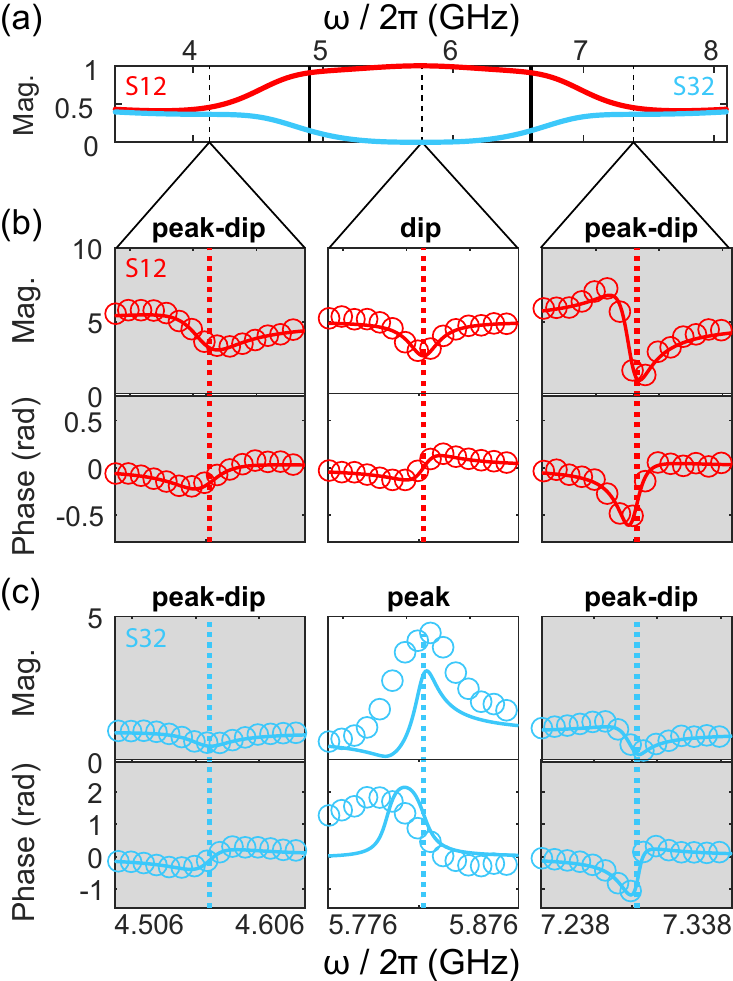}
	\caption{(a) \textit{Cross}-(red) and \textit{through}-(light blue) transmission for the bare interferometer in theory. 
	(b) Measured \textit{cross}-magnitude and \textit{cross}-phase at three different qubit transition frequencies $\omega_{01}/2\pi = \SI{4.556}{\giga\hertz}$, $\SI{5.826}{\giga\hertz}$, and $\SI{7.288}{\giga\hertz}$. 
Left column: peak-dip regime for $\omega_{01}/2\pi\lesssim  \SI{5}{\giga\hertz}$, center column: dip-only regime for $\SI{5}{\giga\hertz} \lesssim \omega_{\textrm{01}} / 2 \pi \lesssim \SI{6.5}{\giga\hertz}$, right column: peak-dip regime for $\omega_{\textrm{01}} / 2 \pi \gtrsim \SI{6.5}{\giga\hertz}$.\protect\\
The solid lines show the results of our transfer matrix model which are in good agreement with the experiment. \protect\\
	(c) Measured \textit{through}-magnitude and \textit{through}-phase at the same qubit transition frequencies: Evidently the dip and peak structure is inverted. In theory the interferometer transmission is zero near the center frequency. This leads to inaccuracies with calibration, and thus, to poor fit quality in the region where the transmitted signal is strongly suppressed.}
	\label{fig:FanoRegimes}
\end{figure}

We probe the qubit in a resonance fluorescence experiment. To this end, we irradiate it with an attenuated coherent microwave signal with an average photon number of much less than one, via one interferometer port and record the magnitude and phase of the through- ($S_{32}$\&$S_{14}$) and the cross- ($S_{12}$\&$S_{34}$) transmission signals as a function of frequency (see figure \ref{fig:IFBSSchematicWMI} for definitions). By varying the magnetic flux threading the dc-SQUID loop of the transmon qubit, we can vary the qubit transition frequency over a wide range and record transmission spectra for different qubit frequencies. Typical spectra are shown in figure\,\ref{fig:FanoRegimes}(b) and (c), where we plot the transmission amplitude and phase as a function of the probe frequency in a narrow frequency windows close to three different qubit transition frequencies $\omega_{01}$.  
In these spectroscopic measurements, the qubit absorbs and reemits microwave photons propagating along the transmission line. 
The presence of the interferometer modifies the interference between the incident and the reemitted signal. 
Depending on the shape of the resonance fluorescence signal, we can identify three different regimes as shown in figure\,\ref{fig:FanoRegimes}: 

\begin{itemize}
	\item peak-dip ($\omega_{\textrm{01}} / 2 \pi \lesssim  \SI{5}{\giga\hertz}$)
	\item dip ($\SI{5}{\giga\hertz} \lesssim \omega_{\textrm{01}} / 2 \pi \lesssim \SI{6.5}{\giga\hertz}$)
	\item peak-dip ($\omega_{\textrm{01}} / 2 \pi \gtrsim \SI{6.5}{\giga\hertz}$)
\end{itemize}

\noindent These regimes can be understood in an intuitive picture: Near its center frequency $\omega_\textrm{IF}$, the interferometer is $\SI{50}{\ohm}$-matched, and therefore, the qubit acts as if it was placed in a bare transmission line\,\cite{Astafiev2010}. 
In this case the presence of the qubit results in the typical symmetric dip in the transmission amplitude due to resonance fluorescence. However, away from the center frequency the interferometer is no longer impedance matched, causing a finite scattering amplitude at the interferometer ports. The resulting constructive and destructive interference effects result in Fano-like shapes of the transmission spectra\,\cite{Fano1961}. 


\subsection{Transfer Matrix Model}\label{sec:model}
The qualitative discussion of the transmission spectra in the previous section shows that a quantitative analysis must take into account all scattering and interference effects on the sample chip. A powerful technique to appropriately model this situation is the transfer matrix approach, where each circuit part is modeled by an individual transfer matrix\,\cite{BornWolf1970}. In our experiments, we measure the transmitted and reflected amplitude of the complete circuit composed of the beam splitter, the transmission lines forming the interferometer arms and the transmon qubit. Therefore, in order to relate the incoming and outgoing modes on one side of the sample with those on the other side in our model approach, we have to construct the total transfer matrix $\mathcal{M}$ representing the whole circuit. This can be done in a straightforward way by modeling each circuit component by its individual scattering matrix (for details, see \ref{ch:transferDetails}). In this way the total matrix $\mathcal{M}$ can be expressed as

\begin{eqnarray}
\mathcal{M}&\equiv&\mathcal{M}(\omega, \omega_{\textrm{IF}},\omega_{\textrm{01}},\Gamma_1,\Gamma_\varphi)\nonumber\\
&=& M_{\textrm{BS}}\cdot M_{\textrm{TL}}\cdot M_{\textrm{Q}}\cdot M_{\textrm{TL}}\cdot M_{\textrm{BS}}\,.
\label{eq:transferMatrix}
\end{eqnarray}

Here, $M_{\textrm{BS}}$, $M_{\textrm{TL}}$, and $M_{\textrm{Q}}$ are the transfer matrices representing the two beam splitters, the transmission lines left and right of the qubit, and the transmon qubit, respectively. Detailed expressions of the different transfer matrices are derived in \ref{ch:transferDetails}. 

The transfer matrix of the transmon qubit is that of a local scatterer. The coefficients of the scattering matrix are determined by the relaxation and dephasing rate as well as the detuning $\delta\omega = \omega - \omega_\textrm{01}$ between the probe signal and the qubit transition frequency. The relaxation and dephasing rates of the qubit can be modeled within the framework of the SBM, describing a single two-level system (TLS) interacting with a bosonic bath. We can apply this model to the transmon qubit interacting with the interferometer, since in the low-probe-power limit the former can be well modeled by an effective TLS, whereas the latter can be described by a broad spectrum of harmonic oscillators\,\cite{pozar}. Since these oscillators are expected to provide memory-less Ohmic damping for the qubit, we expect $\Gamma_1= \alpha \omega_{01}$\,\cite{Leggett1987}. The reason for this linear scaling is the fact that Ohmic dissipation results in an effective bath spectral density which is linear in frequency, $J(\omega) =\beta\omega$ (with high-frequency cutoff $\omega_c$), with $\beta$ corresponding to a friction constant. Note that the dimensionless parameter $\alpha$ reflects the strength of dissipation which in a physical system depends on the amplitude of the noise and its coupling strength. Further details are given in the\,\ref{ap:coupling}. 

Analyzing the measured transmission spectra by the transfer matrix model, we can derive important qubit parameters in a quantitative way.  From a numerical fit of (\ref{eq:transferMatrix}) to the measured cross-transmission data $S_{34}$ and $S_{12}$ [see figure\,\ref{fig:FanoRegimes}(a) and (b)], we extract the qubit transition frequency as well as the relaxation and dephasing rates. Doing so, we discard $S_{14}$ and $S_{32}$, as the almost vanishing signal [see figure\,\ref{fig:FanoRegimes}(a) and (c)] leads to calibration issues and unreliable fits.

Within a $95\%-$confidence interval, the statistical error of the extracted qubit transition frequency is below $2\%$. 
For  $\Gamma_1$ and $\Gamma_\varphi$, we typically observe statistical errors below $33\%$. 
Thus, we obtain a reliable set of data for the decoherence properties of the transmon qubit over a wide range of qubit transition frequencies ($4-8.5\,\si{\giga\hertz}$). 
The excellent fit quality of the individual spectra for most qubit transition frequencies and the reproduction of the regimes described in 
section\,\ref{sec:regimes} (cf. figure\,\ref{fig:FanoRegimes}), provides strong evidence for the validity of the applied transfer matrix model.
Additionally, as expected by design, the interferometer predominantly dictates the transmission spectrum of the system, except in a small region of approximately $100\,\si{\mega\hertz}$ near the qubit transition frequency.
\\ 
\begin{figure*}[!t]
 \centering
        \includegraphics[scale=1]{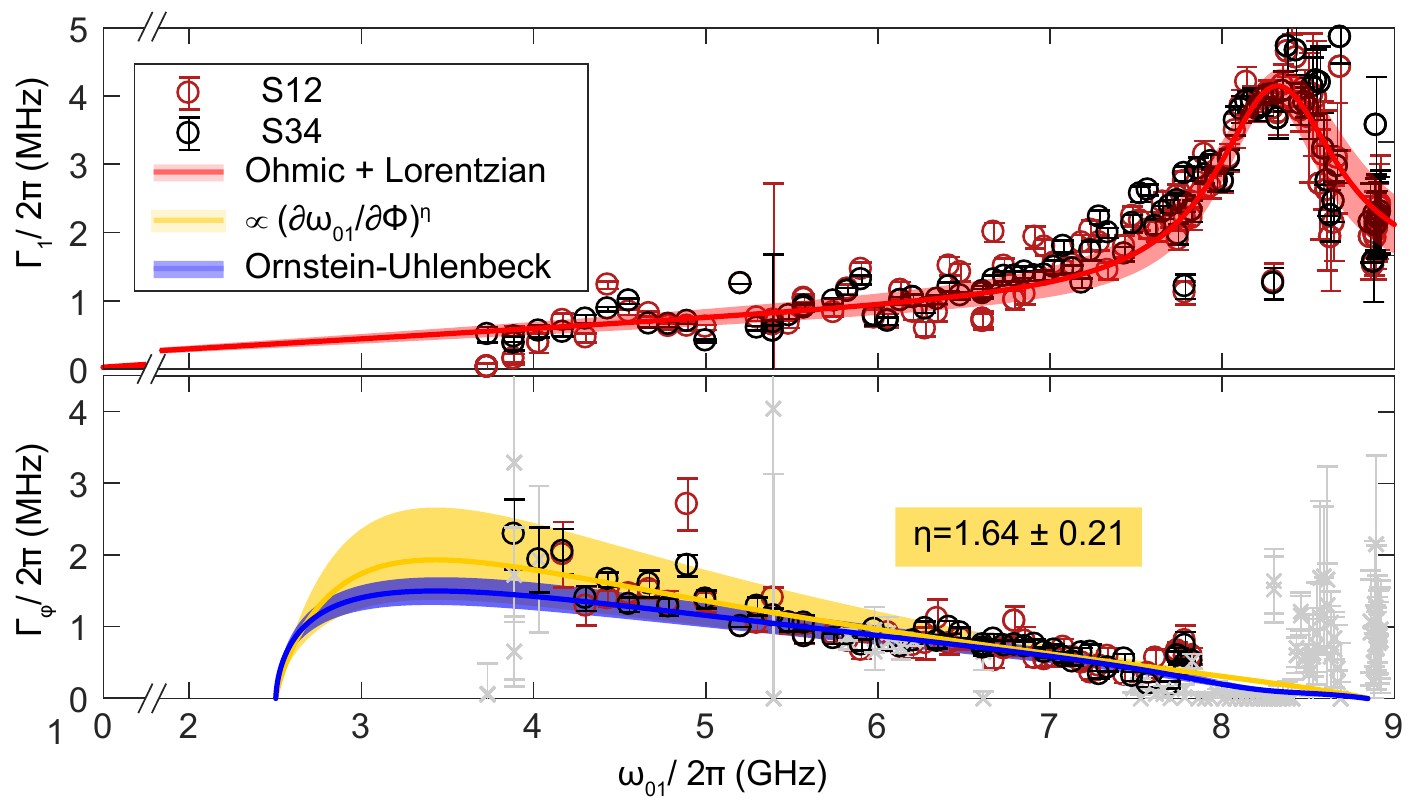}
	\caption{Frequency-dependence of the qubit relaxation and the pure dephasing rate. Measurement data (circles), fit curve (red/yellow/blue lines) with 95\% confidence bounds (shaded areas). Red line: Fit of $\Gamma_1 = \alpha \cdot \omega_\textrm{01} + L(\omega_\textrm{01},\omega_\textrm{L0},\Gamma)$, where $L$ is a Lorentzian with parameters $\omega_\textrm{L0}$ and $\Gamma$. Yellow line: Fit of $\Gamma_\varphi \propto (\partial \omega_{01} / \partial \Phi)^\eta$.	Blue line: Fit of a numerical model based on the Ornstein-Uhlenbeck process to the data.	
For the pure dephasing rate, data points excluded due to large uncertainty (see main text for details) are plotted light grey.}
	\label{fig:Gammas}
\end{figure*}
\subsection{Environment}
\label{sec:ohmic}
In order to gain information about the local electromagnetic environment of the qubit, we use the transmon qubit as a broadband spectrometer in this section.
Following a Golden rule argument, the relaxation rate $\Gamma_1$ is proportional to the noise power spectral density at the qubit transition frequency $S (\omega_{01})$. Hence, the measurement of $\Gamma_1$ as a function of qubit transition frequency allows us to obtain information on $S (\omega)$.

Therefore, we vary the transition frequency of the transmon qubit between $4$ and $\SI{8.5}{\giga\hertz}$ by changing the coil current and, hence, the magnetic flux threading the dc-SQUID loop of the tunable junction. 

From the transmission spectra recorded for every particular qubit frequency we can derive the relaxation and dephasing rates over a wide frequency regime. From this data we can in turn derive valuable information on the interaction of the qubit with its electromagnetic environment. 

More specifically, we perform fits of our transmission model to each recorded spectrum as described in section\,\ref{sec:model} and extract $\omega_\textrm{01}$, $\Gamma_1$, and $\Gamma_\varphi$.
For an Ohmic environment the qubit relaxation rate $\Gamma_1(\omega_\textrm{01})$ is expected to follow $\Gamma_1(\omega_\textrm{01}) = \alpha \cdot  \omega_\textrm{01}$, $\alpha= \tilde{d}^2 / (\hbar c \epsilon_0)$, with the reduced Planck constant $\hbar$, the speed of light $c$,  the electrical vacuum permittivity $\epsilon_0$, and the qubit dipole moment divided by the cross-sectional area of the CPW $\tilde{d}$ (see\,\ref{ap:coupling}). 
Figure\,\ref{fig:Gammas} shows that our $\Gamma_1$ data follows a linear trend for frequencies up to about 7 GHz.
This clearly supports our starting assumption that the transmission line coupled to the qubit provides an Ohmic bath. 
Interestingly, deviations from the Ohmic environment are rather small in the range between $4\,\si{\giga\hertz}$ and $7\,\si{\giga\hertz}$ although the coupling strength is low in comparison to other experiments\,\cite{Haeberlein2015}. For frequencies above $8\,\si{\giga\hertz}$, we observe a pronounced rise in $\Gamma_1$ which
provides a hint to the presence of an additional on-chip mode coupling to the qubit. 
In a first-order approximation, we model this mode by an additional Lorentzian
(center frequency $\omega_\textrm{L0}/2\pi=\SI{8.3}{\giga\hertz}$, full width half maximum $\Gamma/2\pi=\SI{1.5}{\giga\hertz}$) on top of the linear Ohmic background. 
By fitting the data, we find $\alpha = (1.7\pm 0.3)\cdot 10^{-4}$, corresponding to $\tilde{d}=(6.9\pm 2.7)\cdot 10^{-21}\,\si{\ampere\second}$.
In order to find a more quantitative evidence for the transmission line to be the dominant bath for qubit relaxation, we also determine $\alpha_\textrm{res} =\pi (g_\textrm{res}/\omega_\textrm{res})^2=(3.6\pm 0.04)\cdot 10^{-4}$  (see\,\ref{ap:coupling} for details) in the qubit-resonator system used for the precharacterization in section\,\ref{sec:PreCharacterization}.
The good agreement between $\alpha$ and $\alpha_\textrm{res}$ clearly confirms the validity of the SBM-based data analysis.

Finally, we characterize the noise causing pure dephasing of the transmon qubit inside 
the interferometer circuit. 
For the subsequent analysis, we only consider $\Gamma_\varphi$-values with less than $33\%$ 
statistical error (see section\,\ref{sec:model}). 
It is well established that flux noise through the dc-SQUID loop is a dominant source for the fluctuation of the transmon qubit transition frequency $\omega_{01}$, leading to dephasing\,\cite{Ithier2005, Deppe2007, 
Yoshihara2006, Kakuyanagi2007}. As a consequence, we expect a strong dependence of $\Gamma_\varphi(\omega_{01})$ on the first derivative of $\omega_{01}$ with respect to flux $\Phi$ \cite{Deppe2007}.
Indeed, our data is well fitted with the ansatz $\Gamma_\varphi \propto (\partial \omega_{01}/\partial \Phi)^\eta$, as shown in figure\,\ref{fig:Gammas}. Interestingly, the exponent $\eta\simeq 1.64\pm 0.21$ suggests that the observed flux noise may be appreciably correlated rather than simple white noise, for which an exponent of 2 is expected\,\cite{Deppe2007}.
To further characterize the properties of the observed flux noise, we fit the $\Gamma_\varphi(\omega_{01})$ data with a model of Gaussian colored noise (Ornstein-Uhlenbeck process)\,\cite{vanKampen2003, Gardiner1997, Ramos2018}, which allows us to smoothly interpolate between the limits of fast (white) and slow noise. To first order, the transmon transition frequency fluctuates as $\omega_{01}(t)=\omega_{01}+\delta\omega_{01}(t)$, where the deviations $\delta\omega_{01}(t)$ are related to random flux fluctuations via the first derivative as,
\begin{equation}
\delta\omega_{01}(t)=\frac{\partial \omega_{01}}{\partial \Phi}\delta\Phi(t).
\end{equation}
In addition, the colored Gaussian noise model relies on a specific autocorrelation function for the random flux 
fluctuations,
\begin{equation}
\langle \delta\Phi(0) \delta\Phi(\tau)\rangle=\sigma^2e^{-\kappa |\tau|}. 
\end{equation}
Here, $\sigma$ describes the flux noise amplitude and $\kappa$ is a rate describing the temporal range of the correlations  or ``speed of noise''. The noise spectrum corresponding to this model is $S(\omega)\!=\!\!\int e^{i\omega \tau}\langle \delta\omega_{01}(0) \delta\omega_{01}(\tau)\rangle d\tau $.
For $\kappa\rightarrow\infty$, we expect fast noise, because $\lim_{\kappa\rightarrow\infty}S(\omega)=(\partial\omega_{01}/\partial\Phi)^2(2\sigma^2/\kappa)$ becomes constant. The model smoothly connects this white noise limit to the opposite case, $\kappa\rightarrow 0$. Here, one obtains colored quasi-static Gaussian noise, because $\lim_{\kappa\rightarrow 0}S(\omega)\,{=}\,(\partial\omega_{01}/\partial\Phi)^2(2\pi\sigma^2)\delta(\omega)$ diverges for $\omega\,{\rightarrow}\,0$. This limit would correspond to a Gaussian decay envelope in a Ramsey or spin echo type time domain experiment\,\cite{Ithier2005}. 
From a numerical fit of the Ornstein-Uhlenbeck model\,\cite{Ramos2018} to the dephasing data 
$\Gamma_\varphi(\omega_\textrm{01})$, we extract $\sigma=(79\pm 9)\,\mu\Phi_0$ with $\Phi_0$ being the flux quantum. We further find that $\kappa/2\pi$ vanishes within a statistical uncertainty of $\SI{52}{\kilo\hertz}$. Hence, this noise speed is 
negligible with respect to the noise strength $|\partial \omega_{01}/\partial \Phi|(\sigma/2\pi)$, which is on the order of a few megahertz. We conclude that the noise in our device is well described by colored Gaussian noise in the quasi-static limit. This is also consistent with our previous assessment 
based on $\eta\neq 2$ and with a noise spectrum diverging at $\omega/2\pi\simeq \SI{0}{\hertz}$. A possible source for such noise can, e.g., be TLS ensembles produced by surface 
defects in dielectric materials \,\cite{DuttaHorn1981, OMalley2015}.
We can directly relate the quantity sigma to the strength of the $1/f$-noise typically produced by such ensembles \,\cite{Ithier2005,Deppe2007,Kakuyanagi2007,Yoshihara2006}. The standard treatment\,\cite{Ithier2005} provides us with an upper bound  of approximately $100\,\mu\Phi_0$, which is well compatible with the values on the order of a few $\mu\Phi_0$ found in many other works\,\cite{Ithier2005, Deppe2007,VanHarlingen2004, DeppePhDThesis2009}.
\section{Summary}
\label{sec:summary}
We have investigated a complex, engineered on-chip open quantum system consisting of a microwave interferometer with a transmon qubit in 
one of its arms. The interferometer works over a broad frequency range of $\SIrange{4}{8}{\giga\hertz}$ in the microwave domain and the qubit transmission frequency is also tunable over this range. The measured transmission properties are well described by a transfer matrix model, where the scattering matrix of the transmon qubit is derived in the framework of the SBM. 
Using this model, we extract the relaxation rate $\Gamma_1$ and the pure dephasing rate $\Gamma_\phi$ of the qubit 
and discuss their dependence on the qubit transition frequency.

The linear behavior of $\Gamma_1$ with respect to the qubit transition rate $\omega_\textrm{01}$ confirms our expectation that the transmission line acts as an Ohmic bath in the frequency range mentioned above. Above $\SI{8.5}{\giga\hertz}$, our sensitive spectrometer detects a weak enhancement of the spontaneous emission, which we attribute to a spurious on-chip resonance.
We have further analyzed the qubit dephasing rate using a model based on the Ornstein-Uhlenbeck process\,\cite{Ramos2018}.  We find that our circuit QED open quantum system is dominated by slow, colored Gaussian noise. 
Future experiments with higher frequency-resolution or, equivalently, high-resolution time domain experiments, would even allow for the investigation of the microscopic nature of the noise sources.\\
Beyond the current results, our work is also relevant for the analysis of more complex open microwave quantum systems and higher order multi-photon processes\,\cite{Ramos2017}, as they would be required for quantum computing with microwaves in the spirit of all-optical quantum computing\,\cite{Kok2007}. Such experiments would provide a novel approach for QIP, combining established techniques from the optical domain with the advantages of the strong non-linear elements provided by Josephson junctions in the microwave domain. In this context, the present work already implements key elements, such as beam splitters and nonlinearities.

 
\ack
The authors thank all collaborators, especial Edwin Menzel for his efforts in sample fabrication. The authors acknowledge support from the German Research Foundation
through FE 1564/1-1, the IMPRS ’Quantum Science and Technology’, from the MINECO/FEDER Project FIS2015-70856-P, and CAM PRICYT Research Network QUITEMAD+ S2013/ICE-2801. Tomás Ramos further acknowledges the Juan de la Cierva fellowship FJCI-2016-29190. We also acknowledge support from the doctorate program ExQM and the EU project PROMISCE.
\\

\section*{References}
\bibliography{./biblio}
\bibliographystyle{unsrt}
\newpage
\clearpage
\appendix

\section{Details of the cryogenic and measurement setup}\label{ch:setup}
\begin{figure}[!b]
	\centering
	\includegraphics[scale=1]{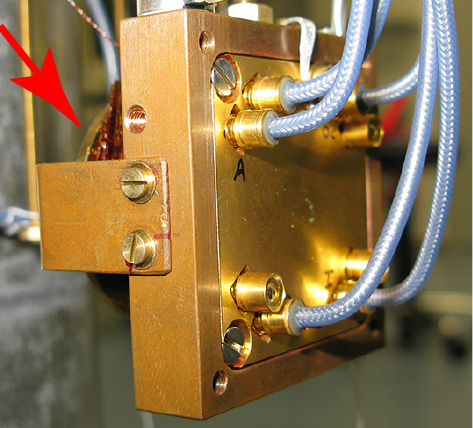}
	\caption{Copper sample box with microwave cabling. A superconducting coil (red arrow) mounted on the backside for tuning the transmon qubit transition frequency.}%
	  \label{fig:SetupPhoto}
\end{figure}
\begin{figure}[htbp]
 \centering
		\includegraphics[scale=1]{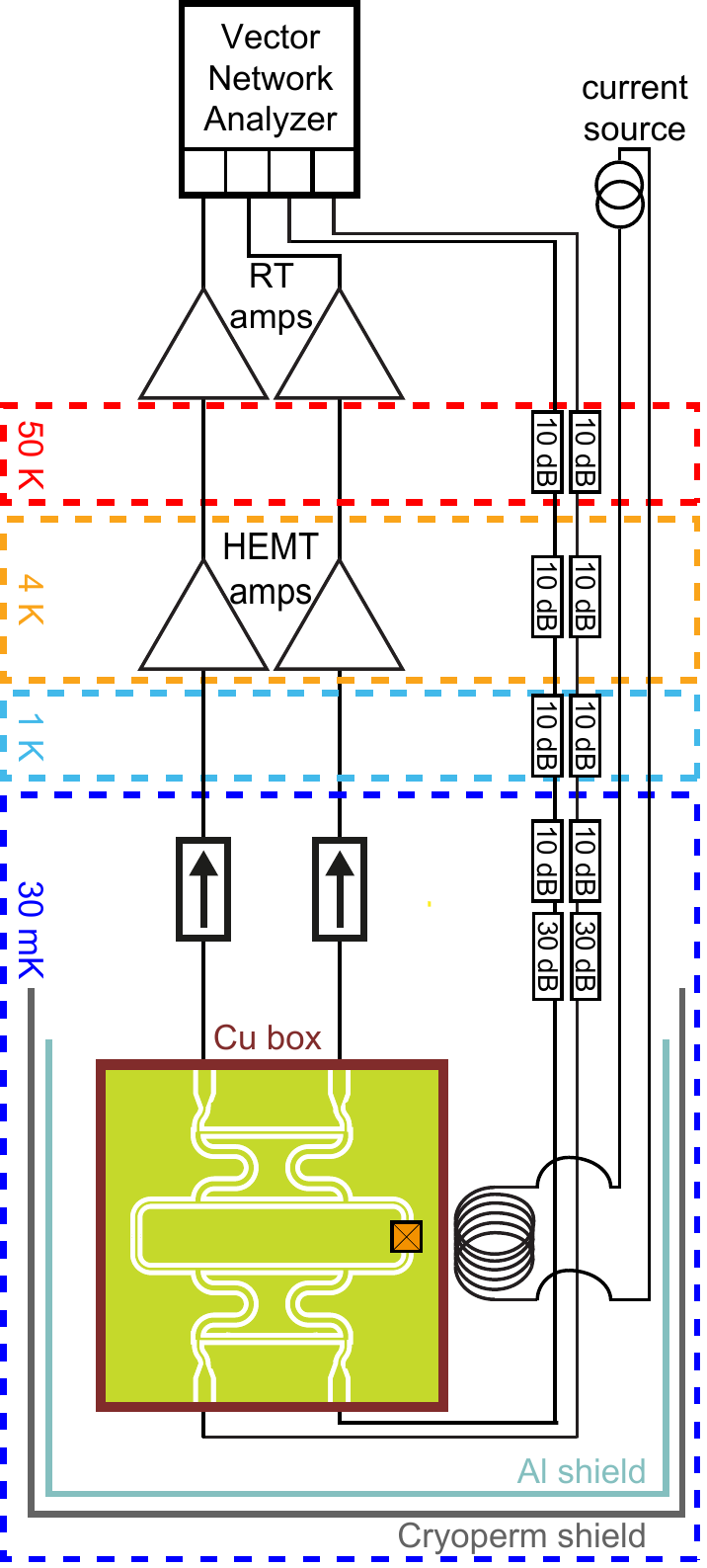}
  \caption{Setup for measurements at millikelvin temperatures. Boxes with an arrow symbolize isolators used to protect the sample from amplifier noise.}
  \label{fig:Setup}
\end{figure}

While the precharacterization of the beam splitter and interferometer samples are done
in a liquid helium bath cryostat, the transmission measurements of the composite system 
are done in a dry dilution refrigerator at a base temperatur of $T\simeq 30 
\si{\milli\kelvin}$. Figure\,\ref{fig:SetupPhoto} shows a photograph of the sample box and the coil and figure\,\ref{fig:Setup} shows a schematic of the setup.

\section{Characteristics in composite system}
The interferometer and transmon qubit properties analyzed in the main text are also 
accessible in the composite system. This means that fitting the transmission 
spectrum at each qubit operating point with our transfer matrix model gives us access to all relevant 
parameters such as the interferometer frequency $\omega_\textrm{IF}$, the qubit transition
frequency $\omega_\textrm{01}$, the relaxation rate $\Gamma_1$, and the pure dephasing
rate $\Gamma_\Phi$.\\

\begin{figure}[t!]
 \centering
\includegraphics[width=.45\textwidth]{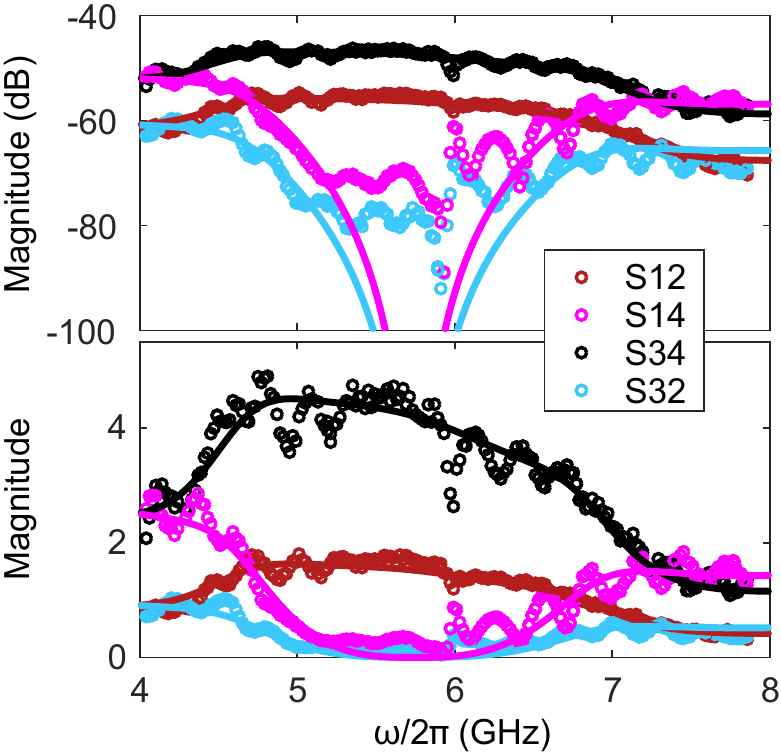}
	\caption{Uncalibrated spectrum of the interferometer. The thin solid lines are fits to the data (symbols) using the transfer matrix model.}
\label{fig:IFdata}
\end{figure}
\begin{figure}[b!]
 \centering
		\includegraphics[scale=1]{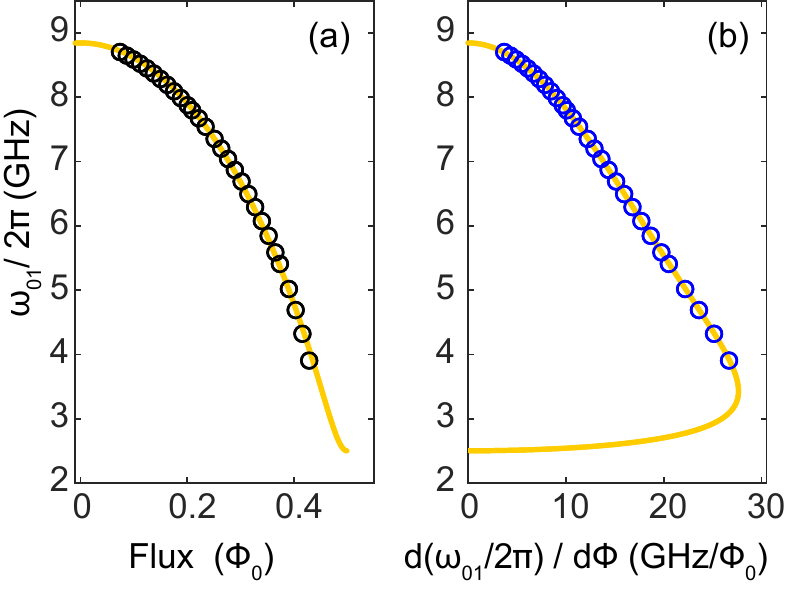}
	\caption{(a) Dependence of qubit transition frequency on the applied magnectic flux in the SQUID loop and (b) its derivative with respect to the applied flux. The yellow lines are fits to the data using the model of the qubit transition-frequency\,\cite{Koch2007}}
\label{fig:fluxdep}
\end{figure}
The transmission spectrum of the interferometer alone is shown in figure\,\ref{fig:IFdata}. It is obtained by measuring the transmission magnitude with the qubit transition frequency tuned outside the accessed frequency range.\\ 


By analyzing the transmission spectra as discussed in section \ref{sec:model} and shown in figure\,\ref{fig:FanoRegimes}, we obtain the flux-dependence of the qubit transition frequency. We fit the theoretically expected behavior\cite{Ithier2005} to the data and obtain $E_C / h=571\,\pm 450\,\si{\mega\hertz}$, and $E_J / h =19.5\,\pm 13\,\si{\giga\hertz}$ (see figure\,\ref{fig:fluxdep}). These values are in good agreement with those measured for the transmon qubit in the transmission line resonator in pre-characterization experiments.


\section{Transfer matrices}

The transfer matrix relates the incoming and outgoing modes $a_\textrm{in}$ and $a_\textrm{out}$ on one side of the scatterer to the outgoing and incoming modes $b_\textrm{in}$ and $b_\textrm{out}$ on the other side. In contrast, a scattering matrix  usually  connects  the  incoming  modes $a_\textrm{in}$ and  $b_\textrm{in}$ on  both  sides  of  the scatterer to the outgoing modes $a_\textrm{out}$ and $b_\textrm{out}$.
%
In the following, we stick to the transfer matrix formalism to model the transmission response of the qubit-interferometer system (see figure\,\ref{fig:Transfermatrix}).
For our sample, the total transfer matrix $\mathcal{M}$ 
connects the complex input signals 
$\left(a_{1,\textrm{in}},\,a_{2,\textrm{in}},\,a_{3,\textrm{in}},\,a_{4,\textrm{in}}\right)$ to the complex output signals 
$\left(a_{4,\textrm{out}},\,a_{3,\textrm{out}},\,a_{2,\textrm{out}},\,a_{1,\textrm{out}}\right)$:
\begin{figure}[!b]
 \centering
	\includegraphics[scale=1]{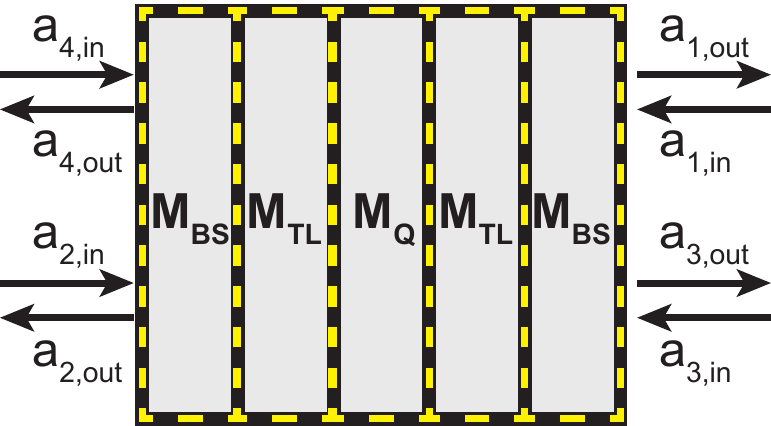}
	\caption{Scheme of the transfer matrix $\mathcal{M}$ for the compete circuit. It defines the relation between incoming modes the 
	$a_{1,\textrm{in}}, a_{2,\textrm{in}}, a_{3,\textrm{in}}$, $a_{4,\textrm{in}}$ and the outgoing modes $a_{1,\textrm{out}}, a_{2,\textrm{out}}, a_{3,\textrm{out}}$ and $a_{4,\textrm{out}}$. Blocks with yellow dotted lines indicate the transfer matrices $M_{\textrm{BS}}$, $M_{\textrm{TL}}$, and $M_{\textrm{Q}}$ of the beam splitter,the transmission lines, and the transmon qubit, respectively.}
	\label{fig:Transfermatrix}
\end{figure}\begin{equation}
  \label{eqn:TransferMatrix}
\left(
\begin{array}{c}
 a_{1,\textrm{out}} \\
 a_{1,\textrm{in}} \\
 a_{3,\textrm{out}} \\
 a_{3,\textrm{in}} \\
\end{array}
\right)
=\mathcal{M}
\left(
\begin{array}{c}
 a_{4,\textrm{in}} \\
 a_{4,\textrm{out}} \\
 a_{2,\textrm{in}} \\
 a_{2,\textrm{out}} \\
\end{array}
\right)
\end{equation}
This property is different from the scattering matrix, which connects the incoming and the outgoing modes at each port. As shown in figure \ref{fig:Transfermatrix}, $\mathcal{M}$ can be decomposed into a product of the matrices $M_{\textrm{BS}}$, $M_{\textrm{TL}}$, and $M_{\textrm{Q}}$ for the beam splitter, transmission line and transmon qubit, respectively. One then obtains (\ref{eq:transferMatrix}).
We set $a_{1,\textrm{in}}=a_{3,\textrm{in}}=0$ assuming that no signal enters the interferometer via port 1 and port 3 on the right side. Technically this assumption is assured by isolators (see figure\,\ref{fig:Setup}).
For the inputs $a_{2,\textrm{in}}$, and  $a_{4,\textrm{in}}$, we assume coherent microwave signals. By solving the system of linear 
equations, we obtain a model function for the different measurement paths illustrated in figure\,\ref{fig:IFBSSchematicWMI}(b).
In the following subsections, we derive expressions for the individual matrices $M_{\textrm{TL}}$, $M_{\textrm{BS}}$, and $M_{\textrm{Q}}$.
\label{ch:transferDetails}
\subsection{Transmission line}
For the transmission line, we take into account the real and imaginary exponent of each signal, thus absorption and phase. This leads to
\begin{eqnarray*}
M_\textrm{TL}=
\left(
\begin{array}{cccc}
e^{i c_1 t}&0&0&0 \\
0&e^{i c_2 t}&0&0\\
0&0&e^{i c_3 t}&0 \\
0&0&0&e^{i c_4 t} \\
\end{array}
\right),\\
\end{eqnarray*}
where $c_i = (i r_i + \phi_i)$. Here, $r_i$ describes the absorption which is treated as a fitting parameter (figure\,\ref{fig:IFdata}). In contrast, $\phi_i$ is the phase which is determined by the length of the transmission line.
\subsection{Beam splitter}
The transfer matrix of the beam splitter provides a $50:50$ splitting of an input signal incident at an input port into the two output ports. It also adds a $180\degree$ phase shift to one of the output signals, resulting in 
\begin{equation*}
M_\textrm{BS}=
\left(
\begin{array}{cccc}
-\frac{i}{\sqrt{2}}		&		0										&		-\frac{1}{\sqrt{2}}  &		0		\\
0											&		\frac{i}{\sqrt{2}}	&		0										&-\frac{1}{\sqrt{2}}		\\
-\frac{1}{\sqrt{2}}		&		0										&		-\frac{i}{\sqrt{2}}  &		0		\\
0											&		-\frac{1}{\sqrt{2}}	&		0										&\frac{i}{\sqrt{2}}		\\
\end{array}
\right)
\end{equation*}

\subsection{Transmon qubit}
The transfer matrix has to reflect the fact that the qubit is placed only in one arm arm of the interferometer. Thus the scattering properties of the qubit have to be taken into account only for one arm of the interferometer, whereas the coefficients for the other arm are just unity:

\begin{equation*}
M_\textrm{Q}=
\left(
\begin{array}{cccc}
-\frac{r^2}{t}	  &		\frac{r}{t}	&	0	&	0 \\
-\frac{r}{t}		&		\frac{1}{t}	&	0	&	0 \\
0		&		0	& 1	& 0 \\
0		&		0	& 0	& 1 \\
\end{array}
\right)
\end{equation*}
Following reference\,\cite{Astafiev2010}, the respective transmission and reflection amplitudes characterizing the scattering by the qubit are given by
\begin{eqnarray*}
\label{eq:tquantum}
r&=&r_0\frac{1-i\left(\delta\omega/\Gamma_{2}\right)}{1+\left(\delta\omega/\Gamma_{2}\right)^2+\Omega^2/\Gamma_{1}\Gamma_{2}},\\
t&=&1-r,\\
\label{eq:rquantum}
\end{eqnarray*}
where $\Omega$ is the qubit Rabi frequency,  
$\Gamma_2\,\equiv\,\Gamma_1/2+\Gamma_\varphi$ is the qubit decoherence rate and $r_0$ is the 
maximum reflection amplitude. 
$\delta\omega\,{\equiv}\,\omega-\omega_\textrm{01}$ is the detuning between the 
drive and the qubit transition frequency. \\

\section{Coupling strength}
\label{ap:coupling}
Here, we briefly derive the coupling strength $g$ of a transmon qubit with transition frequency $\omega_\textrm{01}$ to the Ohmic bath formed by a broadband transmission line in the framework of the SBM. The spin-boson Hamiltonian reads:
\begin{equation*}
	H_\textrm{SBM}=\underbrace{\sum_k \hbar\omega_k \hat{a}_k^\dagger \hat{a}_k}_{H_\textrm{bath}} 
	+\underbrace{\frac{\hbar\omega_\textrm{01}}{2} \hat{\sigma^z}}_{H_\textrm{qb}}
\end{equation*}
\begin{equation*}
	\nonumber  \qquad + \underbrace{\sum_k \left(  g_k\hat{\sigma}^+ \hat{{a_k}} + g_k^*\hat{\sigma}^-\hat{{a_k}}^\dagger \right)}_{H_\textrm{int}},
\end{equation*}
where $H_\textrm{bath}$, $H_\textrm{qb}$ and $H_\textrm{int}$ denote the bath, qubit and interaction part of the Hamiltonian, respectively.\\

The coupling $g_k$ of a qubit to a single photon with frequency $\omega_k$ is given by
a dipole interaction\,\cite{Schoelkopf2008}, and thus 
\begin{equation}
	g_k=\frac{d\cdot E_0}{\hbar}, 
	\label{couplingGk}
\end{equation}
where $d$ is the electric dipole moment of the artificial atom (qubit) and $E_0$ is the root-mean-square electric field due to zero point fluctuations. The electric field contains half of the energy of the transmission line, while the other half is stored in the magnetic field. Therefore,
\begin{equation}
\frac{\epsilon_0}{2}\int  E^2(x)A(x)\, dx=\frac{\epsilon_0}{2}E_0^2 A_0 L=\frac{1}{2}\frac{\hbar \omega_k}{2},
\label{couplingEquality}
\end{equation} 
where $\epsilon_0$ is the vacuum permittivity, $L$ is the length of the waveguide, and $A_0$ is the cross-sectional area perpendicular to the propagation direction. In other words, we determine an effective electric field which is constant inside the mode volume $A_0\cdot L$  and zero elsewhere. 

Solving for $E_0$ in.~(\ref{couplingEquality}), the coupling $g_k$ reads
\begin{equation}
g_k=g_0\sqrt{\frac{\omega_k}{2 L}},\qquad \text{with}\qquad g_0=\sqrt{\frac{\tilde{d}^2}{\hbar\epsilon_0}}.\label{couplingGkResult}
\end{equation}
Here, $\tilde{d}=d/\sqrt{A_0}$ is a qubit dipole moment divided by the cross-sectional area of the CPW $A_0$.
The relaxation rate of the qubit is given in the Markov and rotating wave approximations\,\cite{Diaz2015} by 
\begin{equation}
\Gamma_1=J(\omega_{01}),\label{CouplingRelaxation}
\end{equation}
where the spectral function of the SBM reads,
\begin{equation}
J(\omega)=2\pi \sum_k |g_k|^2 \delta(\omega-\omega_k).\label{CouplingSpectral}
\end{equation}
Substituting\,(\ref{CouplingSpectral}) and\,(\ref{couplingGkResult}) into\,(\ref{CouplingRelaxation}), taking the continuum limit, and assuming a linear dispersion relation $\omega_k=c|k|$, we finally obtain
\begin{equation}
\Gamma_1=\alpha\omega_{01},\qquad \text{with}\qquad \alpha=\frac{g_0^2}{c}.\label{CouplingGammaFinal}
\end{equation}
We see that in this case of an Ohmic bath, the relaxation rate $\Gamma_1$ is proportional to the resonance frequency of the qubit $\omega_{01}$. The authors remark that the $\alpha$ defined here is not the Kondo parameter, also commonly called $\alpha$\,\cite{Haeberlein2015}. If we denote the Kondo parameter as $\alpha_\textrm{K}$, then $\alpha_\textrm{K}=\alpha / (2\pi)$.
%
%
\end{document}